\documentclass{article}
\usepackage{PRIMEarxiv}

\usepackage[utf8]{inputenc} 
\usepackage[T1]{fontenc}    
\usepackage{hyperref}       
\usepackage{url}            
\usepackage{booktabs}       
\usepackage{amsfonts}       
\usepackage{microtype}      
\usepackage{lipsum}
\usepackage{fancyhdr}       
\usepackage{graphicx}       
\usepackage{orcidlink}
\graphicspath{{media/}}     

\title{Modifying Range-Doppler geometry frameworks to process Spotlight SAR imagery in Polar Format 
\thanks{\textit{\underline{Email}}: 
\texttt{piyush.agram@earthdaily.com}} 
}

\author{Piyush S. Agram\, \orcidlink{0000-0003-0711-0264} \\
       EarthDaily Analytics USA Inc.\\
       Minneapolis, MN, USA}

\begin{document}

\maketitle

\begin{abstract}
  We present a simple method to enable processing of Spotlight Synthetic Aperture Radar (SAR) imagery distributed in Polar Format (PFA) using standard Range-Doppler (RDA) geometry algorithms. Our approach is applicable to PFA SAR images characterized by a constant value of the Center of Aperture (COA) time ($t_{COA}$). We present simplified expressions for forward (image-to-ground) and inverse (ground-to-image) geometry mapping using Sensor Independent Complex Data (SICD) conventions~\cite{SICD:design}. We discuss simple changes needed to current open source SAR software that implement Range-Doppler algorithms, to enable support within them for Spotlight data distributed in SICD format. We include a proof-of-concept script that utilizes the Python packages \emph{ sarpy}~\cite{sarpy:lib} and \emph{isce3}~\cite{isce3:lib} to demonstrate the correctness of the proposed approach. 
\end{abstract}

\maketitle

\section{Introduction}

Rapid growth of sub-meter resolution Spotlight image archives acquired by smallsat SAR sensors is an exciting development and a wonderful complement to medium resolution systematic continuous global coverage offered by Sentinel-1. The samples distributed through open data programs~\cite{umbra:opendata, capella:opendata} have further piqued the interest of remote sensing scientists, who are excited to evaluate the suitability of such data for their applications. However, progress has been slow mainly due to the limitations imposed by the formats used to distribute the highest resolution Single Look Complex (SLC) images, for example~\cite{tsai:2024}.

European SAR data providers have offered SLC images in the Zero-Doppler coordinate system~\cite{small:2000} for data acquired in all imaging modes such as stripmap, Spotlight, ScanSAR, TOPS, etc. This is typically achieved by processing all modes of data in a common processor framework, for example~\cite{mittermayer:2003}. This allows end-users to operate on SAR data acquired in different modes without having to switch their analysis tools. ESA's long history of making SAR data publicly available has cemented the Zero-Doppler system as the defacto global community standard and has been adopted by data providers from other countries including Canada, Japan, India, etc.

Smallsat SAR providers from the USA~\cite{umbra:opendata,capella:opendata} have chosen to distribute sub-meter resolution SLCs in Polar Format (PFA) following the Sensor Independent Complex Data~\cite{SICD:design} specification. The SICD standard, developed by the U.S. National Geospatial-Intelligence Agency (NGA), is commonly used by the defense community, but is unfamiliar to the earth observation remote
sensing community. Although commercial SAR software has developed readers for SICD, they do not support the application of end-to-end analytics workflows to SICD data~\cite{tsai:2024}. The absence of public implementations of popular workflows like interferometry and radiometric terrain correction with PFA data is a major deterrent. Further, examples of such workflows often start with freely available data, such as Sentinel-1, in a Zero-Doppler format. It is also interesting to note that Zero-Doppler Spotlight data is already supported by the RGZERO SICD specification~\cite{SICD:design} and the defense community is already able to work with SLC data in that format.

We present a method to adapt Range-Doppler-based SAR processing software to process Spotlight PFA data in SICD format. With our proposed modification, open source SAR processing software can be modified to work with Spotlight PFA data in their original image coordinates. Although it is possible to perform backscatter-related analysis by resampling the PFA imagery onto a Zero-Doppler grid and using existing tools, working in the source image coordinates is paramount for applications that require the phase and spectra information to be preserved, e.g., sub-aperture analytics and interferometric stack generation in native coordinates.

In Section~\ref{sec:sicd}, we simplify the generic SICD PFA geometry model for the specific case of Staring Spotlight data. In Section~\ref{sec:method}, we discuss the implications of our simplified model and present modifications to traditional RDA workflows. The intended audience for this manuscript includes SAR and InSAR engineers and analysts.

\section{SICD Spotlight geometry model}
\label{sec:sicd}
In this section, we walk through SICD's PFA geometry model~\cite{SICD:proj} and derive a simple relationship between pixels in a Spotlight PFA image and Range-Doppler coordinates for the specific case of constant $t_{COA}$ (Staring Spotlight) images. We have examined more than 50 images in the Umbra Open Data Program~\cite{umbra:opendata} and they all share this property. Capella~\cite{capella:opendata} also provides Spotlight data in the SICD PFA format, although most of its public SLC data appear to be in the Zero-Doppler system. We will follow the exact notation used in Section~4.1 of \cite{SICD:proj} to make it easier for the readers to follow.

Since the source PFA image is characterized by constant $t_{COA}$, we can show (Table~\ref{tab:const}) that all the following terms can be treated as constants.

\begin{table}[!h]
\caption{Variables that are constant for constant $t_{COA}$ Spotlight images.}
\begin{center}
\begin{tabular}{|l|l|}
\hline &\\
Variable & Description \\
\hline &\\
$\mbox{ARP}_{COA}^{TGT}$ & Platform position \\
\vspace{0.02cm} &\\
$\mbox{VARP}_{COA}^{TGT}$ & Platform velocity \\
\vspace{0.02cm} &\\
$\left( \mbox{R}^{SCP} \right)_{COA}^{TGT}$ & Range from platform to scene center pixel \\
\vspace{0.02cm} &\\
$\left( \mbox{Rdot}^{SCP} \right)_{COA}^{TGT}$ & Rate of change of range to platform at scene center pixel \\
\vspace{0.02cm} & \\
$\theta_{COA}^{TGT}$ & Polar angle \\
\vspace{0.02cm} & \\
$\left( \frac{d\theta}{dt} \right)_{COA}^{TGT}$ & Rate of change of polar angle w.r.t time \\
\vspace{0.02cm} & \\
$\mbox{KSF}_{COA}^{TGT}$ & Polar aperture scale factor \\
\vspace{0.02cm} & \\
$\left( \frac{d\mbox{KSF}}{d\theta} \right)_{COA}^{TGT}$ & Rate of change of polar aperture scale factor w.r.t polar angle \\
\vspace{0.02cm} & \\
\hline
\end{tabular}
\end{center}
\label{tab:const}
\end{table}

From the material in Section~4.1 of \cite{SICD:proj}, we may write the slant range and the slant range change rate at the image coordinates $(rg^{TGT}, az^{TGT})$ as 

\begin{equation}
\left[ 
\begin{array}{c}
\mbox{R}_{COA}^{TGT} \\
\\
\mbox{Rdot}_{COA}^{TGT}
\end{array} \right] = \left[ 
\begin{array}{c}
\left(\mbox{R}^{SCP}\right)_{COA}^{TGT} \\
\\
\left(\mbox{Rdot}^{SCP}\right)_{COA}^{TGT}
\end{array} \right] + \left[  \begin{array}{cc}
A_{11} & A_{12} \\
\\
A_{21} & A_{22} 
\end{array} \right] \cdot \left[ \begin{array}{c}
rg^{TGT} \\
\\
az^{TGT}
\end{array}\right]
\label{eqn:geommodel}
\end{equation}
where the matrix elements can be written in terms of different constants in Table~\ref{tab:const} as

\begin{eqnarray}
A_{11} & =&  \mbox{KSF}_{COA}^{TGT} \cdot \cos \left( \theta_{COA}^{TGT} \right) \\
A_{12} & = & \mbox{KSF}_{COA}^{TGT} \cdot \sin \left( \theta_{COA}^{TGT} \right) \nonumber \\
A_{21} & = & \left[ \left( \frac{d\mbox{KSF}}{d\theta} \right)_{COA}^{TGT} \cdot \cos \left( \theta_{COA}^{TGT} \right) - \mbox{KSF}_{COA}^{TGT} \cdot \sin \left( \theta_{COA}^{TGT} \right) \right] \cdot \left( \frac{d\theta}{dt} \right)_{COA}^{TGT} \nonumber \\
A_{22} & =&  \left[ \left( \frac{d\mbox{KSF}}{d\theta} \right)_{COA}^{TGT} \cdot \sin \left( \theta_{COA}^{TGT} \right) + \mbox{KSF}_{COA}^{TGT} \cdot \cos \left( \theta_{COA}^{TGT} \right) \right] \cdot \left( \frac{d\theta}{dt} \right)_{COA}^{TGT} \nonumber
\end{eqnarray}
Equation~\ref{eqn:geommodel} shows that the image coordinates of a Spotlight PFA image can be efficiently transformed to Range-Doppler (R-Rdot) coordinates using a simple affine transform. We would note that Doppler frequency ($f_{dop}$) in RDA terminology is related to slant range change rate as 
\begin{equation}
f_{dop} = - \frac{2\cdot \mbox{Rdot}}{\lambda}
\end{equation}
where $\lambda$ is the imaging radar wavelength.

\section{Modifications to standard Range-Doppler algorithms}
\label{sec:method}
In this section, we provide a detailed comparison between forward and inverse mapping methodologies in RDA and PFA frameworks. We present the comparisons in tabular form to assist developers in implementing the presented method within their own RDA frameworks. Here, we simplify the notation from Section~\ref{sec:sicd}, to make the discussion more approachable.

\subsection{Forward mapping}
\label{sec:fwd}
Forward mapping, also known as the image-to-ground operation in SICD~\cite{SICD:proj} terminology, refers to the computation of the location of the target on the ground corresponding to a pixel at range-azimuth indices of $(rg, az)$ in the image raster using a Digital Elevation Model (DEM). Precise implementation of forward mapping is important for transforming ancillary datasets into the SAR image's native grid and aid in interpretation. We provide a detailed one-to-one mapping between the efficient implementation of forward transforms in the Zero-Doppler and PFA frameworks in Table~\ref{tab:forward}.

\begin{table}[!h]
\begin{center}
\caption{Step-by-step comparison of forward mapping computations.}
\label{tab:forward}
\begin{tabular}{|l|l|l|}
\hline & & \\
& Zero-Doppler& Constant $t_{COA}$ PFA \\
\hline & & \\
Constants & 
\begin{minipage}[t]{0.35\textwidth}
\begin{itemize}
\item $R_0$ near range
\item $dR$ range pixel spacing
\item $\mbox{Rdot} = 0$ zero doppler
\end{itemize}
\end{minipage} &
\begin{minipage}[t]{0.35\textwidth}
$\vec{V_{sat}}$, $\vec{R_{sat}}$ platform state vector
\end{minipage}\\
\vspace{0.01cm} & &\\
\hline & &\\
Outer loop & 
\begin{minipage}[t]{0.35\textwidth}
for each azimuth index:
\begin{itemize}
\item Compute $\vec{V_{sat}}$ and $\vec{R_{sat}}$ using orbit information.
\end{itemize}
\end{minipage} & 
\begin{minipage}[t]{0.35\textwidth}
for each azimuth index:
\begin{itemize}
\item $R_0 = R^{SCP} + A_{12} * az$
\item $dR = A_{11}$
\item $\mbox{Rdot}_0 = \mbox{Rdot}^{SCP} + A_{22} * az$
\item $d\mbox{Rdot} = A_{21}$
\end{itemize}
\end{minipage} \\
\vspace{0.01cm} & &\\
\hline & & \\
Inner loop & 
\begin{minipage}[t]{0.35\textwidth}
for each range index:
\begin{itemize}
\item $R = R_{0} + dR * rg$
\end{itemize}
Solve Range-Doppler equation using $\vec{V_{sat}}$, $\vec{R_{sat}}$, $R$, Rdot, DEM.
\end{minipage} & 
\begin{minipage}[t]{0.35\textwidth}
for each range index:
\begin{itemize}
\item $R = R_{0} + dR * rg$
\item $\mbox{Rdot} = \mbox{Rdot}_{0} + d\mbox{Rdot} * rg$
\end{itemize}
Solve Range-Doppler equation using $V_{sat}$, $R_{sat}$, $R$, Rdot, DEM.
\end{minipage}\\
\vspace{0.01cm} & & \\
\hline
\end{tabular}
\end{center}
\end{table}

A closer look at the inner loop for the PFA framework in Table~\ref{tab:forward} shows that both the slant range (R) and the slant range change rate (Rdot) can be represented by an Order-1 polynomial of the range index. This is easily handled by RDA implementations that support Native Doppler (also known as Acquisition Doppler or Non-Zero Doppler)~\cite{isce3:lib,fornaro:2002} geometries. It is often the case that the RDA computations are parallelized for each azimuth index as the range indices share a common satellite state vector ($\vec{R_{sat}}, \vec{V_{sat}}$). In other words, we are able to reuse such an RDA implementation by just including an additional step of updating the near range, slant range pixel spacing and the Doppler polynomial or lookup table for each azimuth line.  

\newpage
\subsection{Inverse mapping}
\label{sec:inv}

Inverse mapping, also known as the ground-to-image operation in SICD~\cite{SICD:proj} terminology, refers to computation of range-azimuth indices in the image raster corresponding to a target ($\vec{T}$) on the ground. Inverse mapping is a faster operation than forward mapping as this involves projection of a target in 3D space to a 2D domain. The primary application of the inverse transform is in geocoding (terrain correction in the SNAP terminology~\cite{snap:lib}).  We provide a detailed one-to-one mapping between the efficient implementation of the inverse transform of a single pixel in the Zero-Doppler and PFA frameworks in Table~\ref{tab:inverse}.

\begin{table}[!h]
\caption{Step-by-step comparison of inverse mapping computation for one pixel}
\begin{center}
\begin{tabular}{|l|l|l|}
\hline & & \\
& Zero-Doppler & Constant $t_{COA}$ PFA \\
\hline & & \\
Constants & 
\begin{minipage}[t]{0.4\textwidth}
\begin{itemize}
\item $R_0$ near range
\item $dR$ range pixel spacing
\item $Rdot = 0$ zero doppler
\end{itemize}
\end{minipage} &
\begin{minipage}[t]{0.35\textwidth}
\begin{itemize}
\item $V_{sat}$, $R_{sat}$ satellite state vectors
\end{itemize}
\end{minipage}\\
\vspace{0.01cm} & &\\
\hline & &\\
Compute & 
\begin{minipage}[t]{0.35\textwidth}
Solve Range-Doppler equation for azimuth time ($\eta$) such that 
$$
\vec{V_{sat}}\left(\eta\right) \cdot \left( \vec{R_{sat}\left(\eta\right)} - \vec{T} \right) = 0
$$

$\left(\left|\vec{R_{sat}\left(\eta\right)} - \vec{T} \right| , \eta \right )$ represent the range-azimuth coordinates in the image.
\end{minipage} & 
\begin{minipage}[t]{0.35\textwidth}
$R = \left| \vec{R_{sat}} - \vec{T} \right|$

$\mbox{Rdot} = \vec{V_{sat}} \cdot \frac{\left(\vec{R_{sat}} - \vec{T}\right)} {\left| \vec{R_{sat}} - \vec{T}\right|}$

Invert Equation~\ref{eqn:geommodel} to get $(rg, az)$ coordinates from $(R, \mbox{Rdot})$.
\end{minipage} \\
\vspace{0.01cm} & &\\
\hline
\end{tabular}
\end{center}
\label{tab:inverse}
\end{table}

We note that all pixels in a PFA image, characterized by constant $t_{COA}$, share the same state vector for the imaging platform ($\vec{R_{sat}}, \vec{V_{sat}}$). This allows us to directly compute the slant range (R) and slant range change rate (Rdot) for the target ($\vec{T}$) as shown in Table~\ref{tab:inverse} and the range-azimuth ($rg, az$) indices using the inverse of Equation~\ref{eqn:geommodel}. Inverse mapping of Spotlight PFA data is based on a simple affine transform and can be implemented using basic numerical computing libraries without specialized radar processing tools. We also note that we can implement a simple PFA to Zero-Doppler resampler by combining the forward mapping for a Zero-Doppler system with the inverse mapping for a PFA system assuming the same reference altitude as the source PFA image.

\subsection{Native-Doppler system}
\label{sec:nativedop}

The SICD framework itself relies on the use of a Native-Doppler solver (Section 5 of \cite{SICD:proj}) to transform (R, Rdot) to ground coordinates. However, the SICD specification does not support imagery in a Native-Doppler system, like the ones distributed by UAVSAR and ALOS-1 providers. The modifications presented in Section~\ref{sec:fwd} would be straightforward in RDA frameworks that already support UAVSAR and ALOS-1 SLC data. If the RDA framework is designed to support only the Zero-Doppler system, for example~\cite{snap:lib}, the framework would need to be generalized to work with the Native-Doppler system before implementing these modifications.

\section{Demonstration}
\label{sec:demo}

A script demonstrating the correctness of the approach presented in this paper can be accessed at \url{https://github.com/piyushrpt/PFA2RDAgeometry}. In the demonstration script, we show that forward and inverse mapping methods via the SICD framework and ISCE's Range-Doppler framework produce the same results (within acceptable numerical error limits). Open-source SAR software developers can use this demonstration script as a reference to implement the methodology laid out in Section~\ref{sec:method} within their RDA frameworks.

\section{Conclusion}

In Section~\ref{sec:sicd}, we show that the geometry model for PFA Spotlight images is significantly simpler than the generic SICD model for PFA data. Building on the simplified model, in Section~\ref{sec:method} we show how RDA geometry algorithms can be adapted to work with PFA Spotlight data. We have also provided a sample implementation using \emph{sarpy} and \emph{isce3} in Section~\ref{sec:demo} to demonstrate the correctness of our method. We hope that open-source SAR processing software developers are able to adapt their Range-Doppler processing frameworks using the proposed approach to allow for more utilization and better exploitation of ever-increasing Spotlight imagery archive being put out in SICD PFA format by smallsat SAR providers.

\section*{Acknowledgments}
Thank you to my colleagues Matthew Calef, Kelly Olsen and Kimberly Carlson for reviewing a draft of this document and providing valuable feedback.

\bibliographystyle{unsrt}
\bibliography{refs}

\begin{thebibliography}{10}

\bibitem{SICD:design}
National~Center for Geospatial Intelligence~Standards.
\newblock Sensor independent complex data (sicd), volume 1, design \& implementation description document.
\newblock Technical Report {NGA.STND.0024-1\_1.2.1}, National Geospatial-Intelligence Agency, Dec 2018.

\bibitem{sarpy:lib}
{National Geospatial-Intelligence Agency (NGA)}.
\newblock Sarpy.
\newblock https://github.com/ngageoint/sarpy.

\bibitem{isce3:lib}
{California Institute of Technology}.
\newblock {InSAR Scientific Computing Environment 3}.
\newblock https://github.com/isce-framework/isce3.

\bibitem{umbra:opendata}
{Umbra Lab Inc.}
\newblock {Umbra Open Data Program}.
\newblock https://registry.opendata.aws/umbra-open-data/.

\bibitem{capella:opendata}
{Capella Space}.
\newblock {Capella Space Synthetic Aperture Radar {(SAR)} Open Dataset}.
\newblock {https://registry.opendata.aws/capella\_opendata/}.

\bibitem{tsai:2024}
Ya-Lun~S Tsai.
\newblock Monitoring summertime erosion patterns over an arctic permafrost coast with recent sub-meter resolution microsatellite sar data.
\newblock {\em EGUsphere}, 2024:1--17, 2024.

\bibitem{small:2000}
David Small and Adrian Schubert.
\newblock Guide to {ASAR} geocoding.
\newblock {\em ESA-ESRIN Technical Note RSL-ASAR-GC-AD}, 1:36, 2008.

\bibitem{mittermayer:2003}
Josef Mittermayer, Richard Lord, and Elke Borner.
\newblock Sliding spotlight {SAR} processing for {TerraSAR-X} using a new formulation of the extended chirp scaling algorithm.
\newblock In {\em IGARSS 2003. 2003 IEEE International Geoscience and Remote Sensing Symposium. Proceedings (IEEE Cat. No. 03CH37477)}, volume~3, pages 1462--1464. IEEE, 2003.

\bibitem{SICD:proj}
National~Center for Geospatial Intelligence~Standards.
\newblock Sensor independent complex data (sicd), volume 3, image projections description document.
\newblock Technical Report {NGA.STND.0024-3\_1.3.0}, National Geospatial-Intelligence Agency, Nov 2021.

\bibitem{fornaro:2002}
Gianfranco Fornaro, Eugenio Sansosti, Riccardo Lanari, and Manlio Tesauro.
\newblock Role of processing geometry in sar raw data focusing.
\newblock {\em IEEE Transactions on Aerospace and Electronic Systems}, 38(2):441--454, 2002.

\bibitem{snap:lib}
{European Space Agency}.
\newblock {SNAP - ESA Sentinel Application Platform v11.0.0}.
\newblock http://step.esa.int.

\end{thebibliography}

\end{document}